Revisiting the measurement of the spin relaxation time in graphene-based devices

H. Idzuchi[1*], A. Fert[2], and Y. Otani[1,3†]

[1] *Center for Emergent Matter Science, RIKEN, 2-1 Hirosawa, Wako 351-0198, Japan*

[2] *Unité Mixte de Physique CNRS/Thales, France associée à l'Université de Paris-Sud, 91405 Orsay, France*

[3] *Institute for Solid State Physics, University of Tokyo, Kashiwa 277-8581, Japan*

[*]idzuchi@riken.jp, [†]yotani@riken.jp

**Abstract**

A long spin relaxation time ($\tau_{sf}$) is the key for the applications of graphene to spintronics but the experimental values of $\tau_{sf}$ have been generally much shorter than expected. We show that the usual determination by the Hanle method underestimates $\tau_{sf}$ if proper account of the spin absorption by contacts is lacking. By revisiting series of experimental results, we find that the corrected $\tau_{sf}$ are longer and less dispersed, which leads to a more unified picture of $\tau_{sf}$ derived from experiments. We also discuss how the correction depends on the parameters of the graphene and contacts.

Spin transport in graphene has strongly attracted attention from the perspective of the long spin relaxation time expected from the small spin-orbit coupling of carbon [1,2]. Spin transport to long distances without spin relaxation, with also the additional interest of the low dissipation of spin currents, is promising for the spintronic devices, in particular to merge functionalities of magnetic non-volatile memory and logic operation, for which a long spin diffusion length $\lambda_{sf}$ enables multi memory elements to function as sources of spin current for input operation via spin-torque switching [3]. However the spin relaxation times $\tau_{sf}$ derived from experiments [4-10], rarely above 1 ns, are much shorter than theoretically expected and also largely dispersed. For example, Volmer *et al.*, see Fig.1(d) in Ref. [9] have clearly highlighted the broad dispersion of $\tau_{sf}$ in wide-ranging series of sample and also pointed out the puzzling general trend of an increase of $\tau_{sf}$ at increasing resistance of the MgO tunnel junction with the magnetic electrodes.

Spin transport in a nonmagnetic conductor NM (metal, semiconductor or graphene) is generally studied [11-17] by using lateral spin valves (LSVs) on which two ferromagnetic (FM) wires are bridged by a nonmagnetic (NM) channel. A spin current is injected from one of the magnetic electrodes and one measures a non-local spin signal related to the spin accumulation in the NM channel. A frequent method to derive the spin relaxation time is to analyze the variation of the spin signal induced by collective spin precession in an applied field, the so-called Hanle effect [16,17]. An important point is that, without a large enough resistance of the contact between channel and electrodes, a part of the injected spin current is reabsorbed by the electrodes (the so-called back-flow current) if the spin resistance $R_N = \lambda_{sf}\rho_N/A_N$, where $\rho_N$ is the electrical resistivity and $A_N$ is the cross section of the NM channel (we first suppose a 3D conductor), is larger than the corresponding spin resistance of the FM electrodes [11-13]. For graphene, most of the determinations of spin relaxation time have been performed by Hanle measurements, and although some numerical simulations have been proposed [8,18] to introduce the spin absorption in the interpretation, almost all the Hanle signal analyses have not taken into account the spin absorption. Here we present analytical expressions [19] of the effect of spin absorption on Hanle curves, and by taking some examples of prior results, we show that, except for the highest contact resistances (tunnel junction resistances), the spin relaxation time had been significantly underestimated. The correction of this underestimate allows us to reduce the dispersion of the spin relaxation times and also to bring closer the results obtained by Hanle measurements and other methods. We can also indicate the values needed for the contact resistances to avoid important spin absorption effects in large ranges of several parameters such as spin diffusion length, spin resistance, and channel length $L$.

First of all, we stress here that the spin absorption is more pronounced in the case of graphene compared to metals. Generally, the spin absorption due to contacts in metallic devices can be easily suppressed with contacts through tunnel junctions. The previous studies by some of us (H. I., Y. O.) clearly showed that, compared to LSVs with metallic contacts, the spin signal $\Delta R_S$ was strongly enhanced by suppression of spin absorption [20] with $Ni_{80}Fe_{20}$(Py)(/MgO)/Ag contacts, as shown in Fig. 1(a) where $\Delta R_S$ increases more than an order of magnitude with varying $R_I$ from 0.1 to 10 Ω. The spin signal $\Delta R_S$ is reduced by the spin absorption if the contact resistance $R_I$ is smaller than the spin resistance $R_N$ of the lateral channel. The different resistance scale of the required contact resistance for metals and graphene comes from the fact that, in typical LSVs with graphene, the channel spin-resistance (~ 10 kΩ) is much larger than with metals ($R_N$ ~ 0.8 Ω for Ag in Fig. 1a). But the fundamental frameworks of spin transport are described in the same manner by replacing the ratio thickness/resistivity of a nonmagnetic metal, $t_N/\rho_N$ by the sheet conductance of graphene, $\sigma_G$ so that the spin resistance of graphene channel is written as $R_N = \lambda_N /(\sigma_G w_N)$ where $\lambda_N$ is the spin diffusion length and $w_N$ is the channel width. The analytical formulation of the absorption effect on Hanle curves was recently established by two of the authors (H. I., Y. O.) and collaborators explaining the reason why the spin transport parameters derived from Hanle curves without taking into account the spin absorption differ from the intrinsic ones [19]. With spin absorption the Hanle voltage is expressed as

$$\frac{V}{I} = -2R_N \left( \frac{P_{F1}}{1-P_{F1}^2} \frac{R_{F1}}{R_N} + \frac{P_{I1}}{1-P_{I1}^2} \frac{R_{I1}}{R_N} \right) \left( \frac{P_{F2}}{1-P_{F2}^2} \frac{R_{F2}}{R_N} + \frac{P_{I2}}{1-P_{I2}^2} \frac{R_{I2}}{R_N} \right) \frac{C_{12}}{\det(\hat{X})}, \quad (1)$$

where $R_{Ik} = 1/G_{Ik}$ ($G_{Ik} = G_{Ik}^\uparrow + G_{Ik}^\downarrow$) and $P_{Ik} = (G_{Ik}^\uparrow - G_{Ik}^\downarrow)/(G_{Ik}^\uparrow + G_{Ik}^\downarrow)$, are the interface resistance (conductance) and the conductance spin asymmetry of $k$-th contact. $R_{Fk} = (\rho_F \lambda_F / A_{Ik})$ and $A_{Fk}$ are the spin resistance and cross-sectional area of the FM electrode on $k$-th contact. The expression $\det(\hat{X})$ is the determinant of the matrix $\hat{X}$ and $C_{12}$ is the (1, 2) component of the cofactors of $\hat{X}$, where $C_{12}$ and $\hat{X}$ are given by

$$C_{12} = -\det \begin{pmatrix} \mathrm{Re}[\bar{\lambda}_\omega e^{-L/\bar{\lambda}_\omega}] & -\mathrm{Im}[\bar{\lambda}_\omega e^{-L/\bar{\lambda}_\omega}] & -\mathrm{Im}[\bar{\lambda}_\omega] \\ \mathrm{Im}[\bar{\lambda}_\omega] & r_{1\perp} + \mathrm{Re}[\bar{\lambda}_\omega] & \mathrm{Re}[\bar{\lambda}_\omega e^{-L/\bar{\lambda}_\omega}] \\ \mathrm{Im}[\bar{\lambda}_\omega e^{-L/\bar{\lambda}_\omega}] & \mathrm{Re}[\bar{\lambda}_\omega e^{-L/\bar{\lambda}_\omega}] & r_{2\perp} + \mathrm{Re}[\bar{\lambda}_\omega] \end{pmatrix}, \quad (2)$$

$$\hat{X} = \begin{pmatrix} r_{1\parallel} + \mathrm{Re}[\bar{\lambda}_\omega] & \mathrm{Re}[\bar{\lambda}_\omega e^{-L/\bar{\lambda}_\omega}] & -\mathrm{Im}[\bar{\lambda}_\omega] & -\mathrm{Im}[\bar{\lambda}_\omega e^{-L/\bar{\lambda}_\omega}] \\ \mathrm{Re}[\bar{\lambda}_\omega e^{-L/\bar{\lambda}_\omega}] & r_{2\parallel} + \mathrm{Re}[\bar{\lambda}_\omega] & -\mathrm{Im}[\bar{\lambda}_\omega e^{-L/\bar{\lambda}_\omega}] & -\mathrm{Im}[\bar{\lambda}_\omega] \\ \mathrm{Im}[\bar{\lambda}_\omega] & \mathrm{Im}[\bar{\lambda}_\omega e^{-L/\bar{\lambda}_\omega}] & r_{1\perp} + \mathrm{Re}[\bar{\lambda}_\omega] & \mathrm{Re}[\bar{\lambda}_\omega e^{-L/\bar{\lambda}_\omega}] \\ \mathrm{Im}[\bar{\lambda}_\omega e^{-L/\bar{\lambda}_\omega}] & \mathrm{Im}[\bar{\lambda}_\omega] & \mathrm{Re}[\bar{\lambda}_\omega e^{-L/\bar{\lambda}_\omega}] & r_{2\perp} + \mathrm{Re}[\bar{\lambda}_\omega] \end{pmatrix}, \quad (3)$$

with $\bar{\lambda}_\omega = \tilde{\lambda}_\omega / \lambda_N$ and $\tilde{\lambda}_\omega = \dfrac{\lambda_N}{\sqrt{1 + i\omega_L \tau_{sf}}}$,

$$r_{k\parallel} = \left( \frac{2}{1-P_{Ik}^2} \frac{R_{Ik}}{R_N} + \frac{2}{1-P_{Fk}^2} \frac{R_{Fk}}{R_N} \right), \text{ and } r_{k\perp} = \frac{1}{R_N G_{Ik}^{\uparrow\downarrow}}, \quad (k=1,2). \quad (4)$$

In the above equation, $G_{Ik}^{\uparrow\downarrow}$ is the spin mixing conductance. In order to reduce the number of fitting parameters, we assume an isotropic spin absorption without additional absorption of the transverse spin components due to spin transfer. Hence the spin mixing conductance is given by $G_{Ik}^{\uparrow\downarrow} = 1/(2R_{Ik} + 2R_{Fk})$. As in the application to LSVs with graphene and Co electrodes ($\lambda_{Co}$ = 38 nm, $\rho_F$ = 25 µΩcm, $P_F$ = 0.36 [22]), the contact $R_I$ is at least three orders of magnitude larger than $R_F$, so that $G_{Ik}^{\uparrow\downarrow}$ is practically equal to $1/(2R_{Ik})$. In the limit of small spin absorption, Eqs. (1)-(4) reduce to the formula used in the interpretation that does not take into account the spin absorption [19]. The inset of Fig. 1(a) for the case of metallic LSVs shows that the contact resistance affects not only the amplitude of spin signal but also the width of the Hanle curves. Actually, Eqs. (1)-(4) enables to explain the different width of Hanle curves in the inset with almost the same spin relaxation time of 40.3±5.3 ps for Py/Ag contacts and 38.0±3.9 ps for Py/MgO/Ag junctions [19].

As shown in Figs. 1(c)-(e), we have reanalyzed the Hanle curves of Ref. [4] for graphene-based LSVs with contact resistances ranging between 0.285 kΩ and 30 kΩ sample. All the data are taken at room temperature for graphene of typical conductivity around 0.4 mS and in samples with usual values of $w_N$ and $L$ (Length) in the µm range. For the sample of Fig. 1c ($R_I$ = 0.285 kΩ), a good fit of the curve is obtained by using the parameters of the first line in Table I, i.e. with the experimental parameters of Ref.

[4], and, for the three free parameters, $\tau_{sf}$ = 498 ps, $P_I$ = 0.0108, $D_N$ = 0.0149 m$^2$/s. It should be noted here that $\tau_{sf}$ is considerably increased with respect to the value derived without considerin

g the spin absorption, $\tau_{sf}^*$ = 84 ps. Therefore in order to characterize intrinsic spin transport properties, it is indispensable to consider the spin absorption which significantly modifies the spatial distribution of the chemical potential inside the nonmagnetic channel [19].

For the sample of Fig. 1(d) with $R_I$ = 6 k$\Omega$ (characterized by Han *et al.* as a sample with pinhole in the tunnel junctions) the spin relaxation time we obtained with the parameters of Table I is 359 ps, whereas $\tau_{sf}^*$ = 134 ps had been obtained without spin absorption [4]. For the sample with a mean value of the tunnel contact resistance respectively equal to 30 k$\Omega$ and 50 k$\Omega$ in Ref. [4], our analysis of the Hanle curve (see Fig. 1(e) for one of the samples), leads respectively to $\tau_{sf}$ = 481 ps and $\tau_{sf}$ = 511 ps whereas the standard analysis [4] had led to $\tau_{sf}^*$ = 448 ps and 495 ps.

All the data are summarized in Fig. 2 by the plot of $\tau_{sf}$ and $\tau_{sf}^*$ as a function of the interface resistance for all the sample in Ref. [4]. Whereas the original analysis leads to an apparent continuous contact-induced increase of $\tau_{sf}^*$ from 84 ps for $R_I$ = 0.285 k$\Omega$ to 495 ps for $R_I$ = 50 k$\Omega$, we find that, with contact absorption, all the Hanle curves can be accounted for with an almost constant "intrinsic" $\tau_{sf}$ of about 500 ps and an almost constant $\lambda_{sf}$ close to 3 µm. The increase of spin relaxation time when the spin absorption is taken into account is still by a factor of 2.7 for the sample with relatively close values of $R_I$ and $R_N$ (6 kΩ and 9.11 kΩ, respectively). The correction factor still amounts to 7% (3%) for the interface resistance as large as 30 kΩ (50 kΩ).

An apparent increase of spin relaxation time with the contact resistance similar to that in Ref. [4] was also reported by Volmer *et al.*[9]. We could interpret also these results by taking into account the effect of spin absorption on the Hanle curves and explain the general trend of the interface with a single value of $\tau_{sf}$ for each sample series of single-layer and bilayer graphene (not presented here).

We want now to discuss how, in our model, a precise determination of $\tau_{sf}$ can be affected by the uncertainty of the other parameters. The polarization $P_I$ determines the amplitude of the spin signal but affects only very weakly its relative variation in Hanle curves. It can be seen from Eqs. (1)-(4) that $P_I$ is in the field-independent first factors but influences only very weekly the factor $C_{12}$ governing the field dependence of Hanle signal (it can be easily seen from the expected small influence of $P_I$ via $r_{k\parallel}$ in Eq.(4)). It can be consequently said that some uncertainty on $P_I$ and the dispersion of its values affect very weakly our determination of $\tau_{sf}$. There is also some uncertainty on the parameter $\sigma_G$ entering the calculation. For example, when $\sigma_G$ = 0.35 mS instead of 0.44 mS, $\tau_{sf}$ becomes 521 ps instead of 498 ps for the sample in the first row on Table I. Similarly, when $w_N$ = 2.2 μm instead of 1.0 μm, $\tau_{sf}$ becomes 397 ps instead of 498 ps because of the decrease in the spin absorption as $R_I/R_N$ ($\propto w_N$) increases. However these variations are small compared to the total correction.

Finally, after having illustrated the influence of the spin absorption by the contacts in the specific case of the sample in Ref. [4], we will describe how, more generally, the effect of the contacts varies in different parameter ranges. In Fig. 3(a) we show the correction factor $\tau_{sf}/\tau_{sf}^*$ as a function of $R_I/R_N$ for several values of the ratio $L/\lambda_N$ (the other parameters are typical for graphene: $\sigma_G$ = 0.335 mS, $P_I$ = 0.1 , $w_N$ = 1.0 μm, $D_N$ = 0.010 m$^2$/s). The Hanle curves are first calculated using our model with spin absorption for a series of values for $\tau_{sf}$ (which give the corresponding values of $\lambda_N$ and $R_N$). Then, in a second stage, the "non-corrected relaxation time" $\tau_{sf}^*$ is derived from fitting these curves to the standard expression without spin absorption [17]. It turns out that the correction factor becomes very large when $L$ is shorter than $\lambda_N$. With $L/\lambda_N$ = 0.1, $\tau_{sf}/\tau_{sf}^*$ is about 10 for $R_I = R_N$ and is still 1.6 for $R_I = 10\ R_N$. This means that, for example with a sample of graphene of sheet resistivity 1 kΩ and $\lambda_N$ = 10 μm, and a LSV with $L$ = 1 μm and $w_N$ = 1 μm (resulting in $R_N$ = 10 kΩ), the correction factor is still as large as 1.6 for $R_I = 10R_N = 100$ kΩ, the resistance larger than that of most tunnel resistance used up to now in graphene LSVs.

In Fig. 3(b), the solid lines indicate how, for different values of $R_I$ and with the same value of $\sigma_G$, $P_I$, $w_N$, and $D_N$ (=0.010m$^2$/s) as in Fig. 3(a), the $\tau_{sf}$* derived from a non-corrected interpretation of Hanle curves [17] varies as a function of the $\tau_{sf}$ used to calculate these curves. The dashed and dotted lines are for the single value $R_I = 100$ kΩ and two different values of $D_N$ (=0.050m$^2$/s and 0.0020m$^2$/s). It turns out that, when the intrinsic spin relaxation time $\tau_{sf}$ is long, the $\tau_{sf}$* is more strongly underestimated, much more than in our above re-interpretation of the situation with relaxation times below 1 ns in Ref. [4]. Typically, the results of Fig. 3(b) show that, even with a resistance of 100 kΩ, a spin relaxation time of 5 ns is underestimated by a factor of 2.5 when $D_N = 0.05$ m$^2$/s [23].

In summary, we have examined the effect of spin absorption on the determination of the spin relaxation time in graphene from Hanle experiments. We have revisited Hanle curves of graphene LSVs which were previously analyzed by using the "standard" model without spin absorption [17]. Our reanalysis shows that the reported difference in the spin relaxation times of samples with different contact resistances is due to interface effects with the contacts. After correction of these effects the spin relaxation times are much less dispersed. A general discussion based on the results of Fig. 3 shows that, without correction for the back flow and the spin absorption through contacts, and even with contact resistances as large as 100 k$\Omega$, the spin relaxation time is significantly underestimated when its intrinsic value is in the range of a few nanoseconds, especially when the graphene length is smaller than its spin diffusion length, or when, in highly conductive graphene, the diffusion constant $D_N$ amounts to few 0.01 m$^2$/s. The characterization of intrinsic spin relaxation time taking account of the effect of spin absorption on the Hanle curves give a hint to interpret recent results such as similar spin relaxation time for suspended and non-suspended graphene the latter of which has an additional contact [24] and the difference between the spin relaxation derived from Hanle measurements and the slower spin relaxation obtained by other approaches [7].

## References and footnotes

**Figure captions**

**FIG. 1.** (a) Non-local spin signal $\Delta R_S$ as a function of contact resistance $R_I$ for NiFe/Ag/NiFe lateral spin valves (LSVs) with MgO layer between NiFe(Py) and Ag, for the separation $L$ = 0.30 µm [20]. The crossover resistance between the conductivity mismatch regime (too fast spin absorption by the contacts) and saturation (intrinsic spin relaxation in Ag), $R_{crossover} = R_N = \rho_{Ag}\lambda_{Ag}/t_{Ag}w_{Ag}$ is the scale governing the variation with $R_I$ [21] and is of the order of 1 Ω for a metal as Ag (resistivity $\rho_{Ag}$ in the µΩcm range, spin diffusion length $\lambda_{Ag}$ around 1 µm and the thickness $t_{Ag}$ (width $w_{Ag}$) in the $10^1$nm ($10^2$nm) range. Inset: Example of the different widths of Hanle curves for two LSVs with different contact resistances. Taking into account the spin absorption enables a fit of the Hanle curves of the two samples (see solid lines) with practically the same spin relaxation time. (b) Schematic illustration of Hanle measurement. Larmor precession of spin current is observed electrically. (c) Fit of Hanle curves for single layer graphene lateral spin valves with transparent contacts ($R_I$ = 0.285 kΩ), (d) with pinhole in the tunnel contacts ($R_I$ = 6 kΩ), and (e) with tunnel junctions ($R_I$ = 30 kΩ). $\tau_{sf}$ is the intrinsic spin relaxation time derived from the model with the effect of spin absorption whereas the spin relaxation time $\tau_{sf}*$ is derived without taking into account the spin absorption ("standard" model) in Ref. [4].

**FIG. 2.** Comparison of the spin relaxation time $\tau_{sf}$ (circles) and $\tau_{sf}^*$ (triangles) respectively derived in models with (this paper) and without [4] spin absorption for samples of various contact resistance $R_I$. Solid lines are guide to the eyes. Dashed line is the approximate trend in a similar plot of spin relaxation times derived with taking no account of spin absorption in Ref. [9].

**FIG. 3.** (a) General variation of the spin absorption correction factor of the spin relaxation time, $\tau_{sf}/\tau_{sf}^*$, as a function of $R_I/R_N$ for several values of $L/\lambda_N$. The correction factor is large for $R_I/R_N < 1$ (strong spin absorption) but it can also be significant for values of $R_I$ as large as 5-10 $R_N$ if the channel length is shorter than the spin diffusion length. Curves were calculated with typical parameter for graphene: $\sigma_G$ = 0.335 mS, $P_I$ = 0.1, $w_N$ = 1.0 µm, $D$ = 0.010 m²/s. (b) Solid lines: uncorrected $\tau_{sf}^*$ vs intrinsic $\tau_{sf}$ for several values of $R_I$ calculated for graphene LSVs with, in addition to the parameters used in (a), $L$ = 1 µm and a typical value for $D_N$ = 0.010 m²/s. The dotted (dashed) lines are calculated for the only value $R_I$ = 100 kΩ with $D_N$ = 0.050 m²/s ($D_N$ = 0.0020 m²/s) to show that the correction becomes more important for high mobility graphene (for example $D_N$ = 0.050 m²/s in recent experiments [23]). D and R in legend represent $D_N$ and the contact resistance $R_I$, respectively. One sees that even a contact resistance of 100 kΩ is not enough large to block spin absorption when $\tau_{sf}$ or $D_N$ are large. Typical value for $\lambda_N$ is given by $\lambda_N = (D_N \tau_{sf})^{1/2}$ = 5 µm for $D_N$ = 0.01 m²/s and $\tau_{sf}$ = 2.5 ns.

**Tables**

Table I: Parameters for the interpretation of Hanle signals in Figs. 1(c)-(e). $R_I$, $\sigma_G$, $L$, $w_N$ are from Ref. [4], $D_N$, $P_I$ and $\tau_{sf}$ are the free parameters ($\lambda_N$ and $R_N$ are the corresponding values of the spin diffusion length and spin resistance), and $\tau_{sf}*$ is the spin relaxation in the previous interpretation [4] in a model without spin absorption. $P_I$ shown with † was the geometric mean of $P_I$ for Hanle curves with parallel and antiparallel magnetic configurations.

| Junction | $R_I$ (kΩ) | $\sigma_G$ (mS) | $L$ (μm) | $w_N$ (μm) | $D_N$ (m²/s) | $P_I$ | $\tau_{sf}$ (ps) | $\lambda_N$ (μm) | $R_N$ (kΩ) | $\tau_{sf}*$ (ps) |
|---|---|---|---|---|---|---|---|---|---|---|
| Transparent | 0.285 | 0.44 | 3.00 | 1.00 | 0.0149 | 0.0108 | 498 | 2.72 | 6.18 | 84 |
| Pinhole | 6.00 | 0.27 | 2.00 | 1.00 | 0.0168 | 0.120† | 359 | 2.46 | 9.11 | 134 |
| Tunnel | 30.0 | 0.29 | 5.50 | 2.20 | 0.0134 | 0.0810 | 481 | 2.54 | 4.26 | 448 |
| Tunnel | 50.0 | 0.29 | 2.10 | 2.20 | 0.0176 | 0.156† | 511 | 3.00 | 4.70 | 495 |

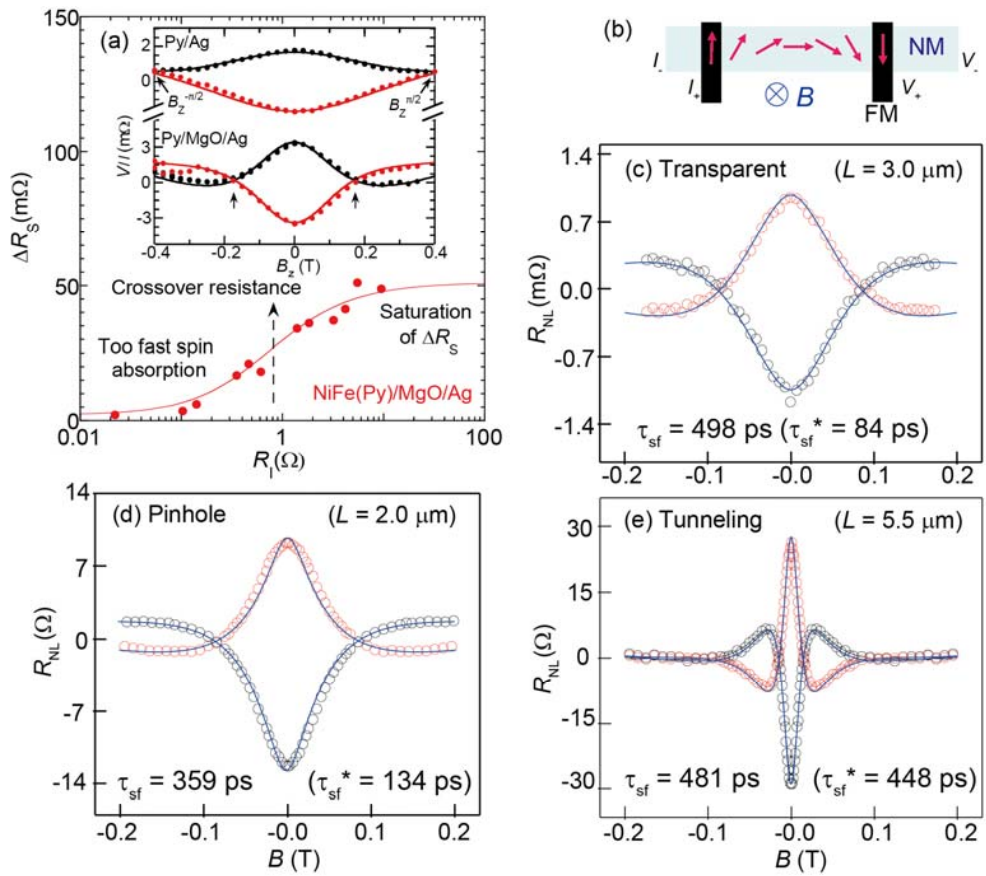

Fig.1 Idzuchi *et al*.

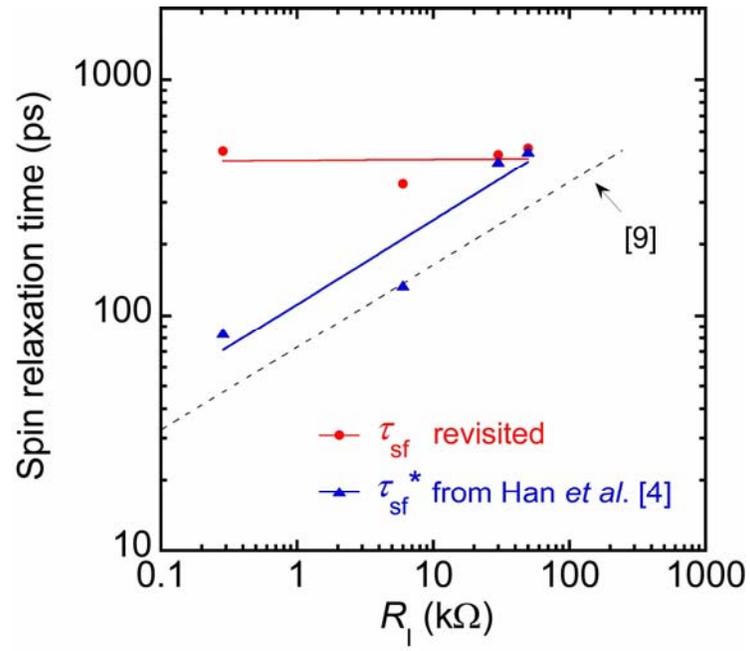

Fig.2   Idzuchi *et al.*

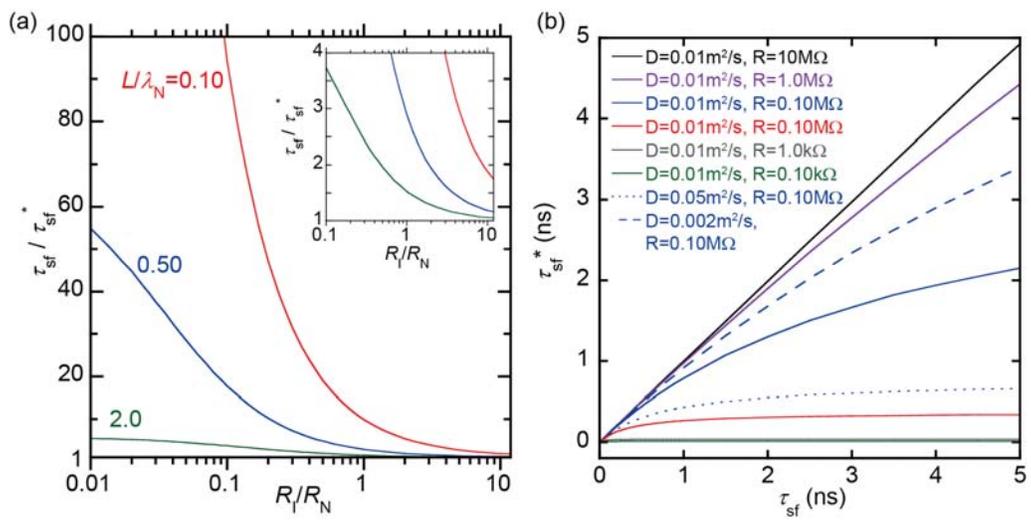

Fig.3 Idzuchi *et al*.